\documentclass[final,5p,times,twocolumn]{elsarticle}

\IfFileExists{srcltx.sty}{\usepackage[inactive]{srcltx}}

\usepackage{graphicx}
\usepackage{amssymb}
\usepackage{lineno}

\usepackage{url}

\biboptions{sort&compress}

\newcommand\araa{ARA\&A}%  % Annual Review of Astron and Astrophys 
\newcommand\apj{ApJ}%    % Astrophysical Journal 
\newcommand\apjl{{ApJ}}%     % Astrophysical Journal, Letters 
\newcommand\aap{{A\&A}}%     % Astronomy and Astrophysics 
\newcommand\jcap{{J. Cosmology Astropart. Phys.}}% Journal of Cosmology and Astroparticle Physics
\newcommand\mnras{{MNRAS}}%   % Monthly Notices of the RAS 
\newcommand\prd{{Phys. Rev.~D}}% % Physical Review D 
\newcommand\pasj{{PASJ}}%     % Publications of the PASJ 
\journal{Astroparticle Physics}

\begin{document}

\begin{frontmatter}


\title{Prospects for future very high-energy gamma-ray sky survey: impact 
of secondary gamma rays}


\author[kipac]{Yoshiyuki Inoue}
\ead{yinoue@slac.stanford.edu}

\author[inr]{Oleg E. Kalashev}
\ead{kalashev@inr.ac.ru}

\author[ucla,ipmu]{Alexander Kusenko}
\ead{kusenko@ucla.edu}

\address[kipac]{Kavli Institute for Particle Astrophysics and Cosmology, Department of Physics, Stanford University and SLAC National Accelerator Laboratory, 2575 Sand Hill Road, Menlo Park, CA 94025, USA}
\address[inr]{Institute for Nuclear Research, 60th October Anniversary Prospect 7a, Moscow 117312 Russia}
\address[ucla]{Department of Physics and Astronomy, University of California, Los Angeles, CA 90095-1547, USA}
\address[ipmu]{Kavli IPMU (WPI), University of Tokyo, Kashiwa, Chiba 277-8568, Japan}

\begin{abstract}
Very high-energy gamma-ray measurements of {\em distant} blazars can be well explained by secondary gamma rays emitted by cascades induced by ultra-high-energy cosmic rays. The secondary gamma rays will enable one to detect a large number of blazars with future ground based gamma-ray telescopes such as Cherenkov Telescope Array (CTA). We show that the secondary emission process will allow CTA to detect 100, 130, 150, 87, and 8 blazars above 30 GeV, 100 GeV, 300 GeV, 1 TeV, and 10 TeV, respectively, up to $z\sim8$ assuming the intergalactic magnetic field (IGMF) strength $B=10^{-17}$~G and an unbiased all sky survey with 0.5 hr exposure at each Field of View, where total observing time is $\sim540$ hr. These numbers will be 79, 96, 110, 63, and 6 up to $z\sim5$ in the case of $B=10^{-15}$~G. This large statistics of sources will be a clear evidence of the secondary gamma-ray scenarios and a new key to studying the IGMF statistically. We also find that a wider and shallower survey is favored to detect more and higher redshift sources even if we take into account secondary gamma rays.
\end{abstract}


\end{frontmatter}

%\linenumbers

\section{Introduction}
\label{sec:intro}

Current imaging atmospheric Cherenkov Telescopes (IACTs) have already found $\sim$140 very high-energy (VHE; above 100~GeV) gamma-ray sources, including $\sim$50 blazars up to redshift $z\sim0.5$ \footnote{\url{http://tevcat.uchicago.edu}}. Very recently Furniss et al.~\cite{fur13} reported the redshift lower limit of $z>0.6035$ for the VHE gamma-ray blazar PKS 1424+240 and Tanaka et al.~\cite{tan13} reported detection of two VHE gamma-ray photons from the blazar PKS 0426-380 at $z=1.1$ using the Fermi gamma-ray telescope ({\it Fermi}).  Blazars, a class of active galactic nuclei (AGNs), are the dominant population in the extragalactic gamma-ray sky. VHE gamma rays propagating through intergalactic space are attenuated by photon-photon pair production interaction ($\gamma\gamma\rightarrow e^+e^-$) with photons of the extragalactic background light (EBL) from far-infrared to ultraviolet wavelengths, see, e.g., Refs.~\cite{gou66,jel66,ste92,dwe13}. Recent studies have detected attenuation of gamma rays on 
EBL~\cite{2012Sci...338.1190A,2013A&A...550A...4H,2013A&A...554A..75S}, 
using a dataset dominated mostly by optical depth of the order of 
1. However, distant blazars appear to have harder spectra than one would expect from simple gamma-ray emission models~\cite{ste92}, as well as a redshift dependence of the observed spectral index that is different from what was expected~\cite{Stecker:2006wq,ess12}, although a large uncertainty remains in the measured redshifts and spectral indeces~\cite{Sanchez:2013lla}.  This has prompted a number of exotic scenarios based on  
hypothetical axion-like particles \cite{dea07,sim08,san09} (different from the QCD axion), as well as  Lorentz invariance violation \cite{kif99,pro00}  as possible explanations of the spectral hardening.  

An alternative interpretation is the cascade emission from high energy protons propagating through intergalactic space. Since AGN jets are believed to be powerful sources of cosmic rays, protons generated in AGN jets can interact with the EBL via the reactions $p\gamma\rightarrow p\pi^0$ and $p\gamma\rightarrow n\pi^+$. Pions quickly decay into high energy photons and electrons. Furthermore, the high-energy protons can interact with the cosmic microwave background (CMB) photons via $p\gamma\rightarrow p e^+e^-$.  Both of these channels initiate electromagnetic cascades distributed uniformly along the line of sight, and the highest energy gamma rays produced relatively close to the observer are not attenuated significantly by the EBL.   Interactions of high-energy protons with the photon backgrounds along the line of sight can produce point images of sources observable by IACTs, as long as the intergalactic magnetic fields deep in the voids are in the femtogauss range~\cite{ess_magnetic}. 
The observed fluxes are comprised of two components: primary gamma-ray flux produced at the source or in the subsequent electromagnetic cascade (not including any cosmic ray interactions), and secondary gamma-ray flux, which arises from line-of-sight interactions of cosmic rays. 
This secondary gamma-ray scenario can reproduce the observed spectra of distant blazars  remarkably well~\cite{ess10,ess10_prl,ess11,Razzaque:2011jc,mur12,aha13,2013ApJ...764..113Z,tak13}. 

It is expected that the next generation IACT, Cherenkov Telescope Array (CTA)~\cite{act11,ach13}, will be able to detect a large number of blazars with an extragalactic blank field sky survey \cite{ino10,dub13}. A statistical study of VHE blazars in the CTA era will provide a crucial key to understanding of AGN populations and high-energy phenomena around supermassive black holes in AGNs. However, the expected number of blazars to be observed above $1$~TeV is limited by both the observational time and the interactions with the EBL~\cite{ino10,dub13}. The purpose of this paper is to consider the impact of the secondary gamma rays on future blazar surveys by CTA \citep[see also][]{mur12_cas}, since secondary gamma rays avoid significant EBL attenuation at energies above $1$~TeV \cite{ess11,ess12}. 

For this purpose, the blazar gamma-ray luminosity function (GLF), primary spectral energy distribution (SED), and secondary gamma-ray SED are needed. The blazar GLF has been studied in detail in many papers~\cite{nar06,ino09,ino10,aje12,har12}. In Refs.~\cite{ino09,ino10,har12} blazar GLFs were constructed taking into account the blazar SED sequence~\cite{fos98,kub98,don01}, in which the synchrotron and inverse Compton (IC) peak photon energies decrease as the bolometric luminosity increases. These models are in good  agreement with the {\it Fermi} data, and they allow one to predict blazar evolution at any wavelength by incorporating the GLF with SED. In this paper, we use the model of Ref.~\cite{ino10} to predict the expected number and redshift distributions of VHE blazars in future CTA blank field sky surveys. We adopt secondary gamma-ray spectral models of Ref.~\cite{kal13} with the EBL model of \cite{ino13} assuming the range of parameters that can explain the known hard TeV blazars.

This paper is organized as follows. We introduce GLF models and blazar SEDs, as well as the model of VHE gamma-ray absorptions by EBL in section 2. In section 3, we study the impact of the secondary gamma rays on future VHE sky surveys assuming certain observing modes. Our results are summarized in section 4. Throughout this paper, we adopt the standard values of cosmological parameters $(h, \Omega_M, \Omega_\Lambda) = (0.7, 0.3, 0.7)$.

\section{Model Description}
\label{sec:mod}

\subsection{Blazar Gamma-ray Luminosity Function and Primary Spectrum}
In Refs.~\cite{ino09} and \cite{ino10} a blazar GLF model was developed for luminosity at 100~MeV $L_{\gamma, 100 \rm MeV}$  based on the latest determination of X-ray luminosity function of AGNs \cite{ued03,has05}, featuring the so called luminosity dependent density evolution. Another new feature was taking into account the blazar SED sequence. Blazar sequence is a feature seen in the mean SED of blazars that the synchrotron and IC peak photon energies decrease as the bolometric luminosity increases \cite{fos98,kub98,don01,ghi09}. The key parameters in GLF have carefully been determined to match the observed flux and redshift distribution of EGRET blazars by a likelihood analysis. The predicted extragalactic gamma-ray background (EGB) spectrum~\cite{ino10} including contributions from Seyferts~\cite{ino08} and radio galaxies~\cite{ino11} are in good agreement with the EGB spectrum reported by {\it Fermi}~\cite{abd10_egb, ino12_proc}. Predicted blazar GLF~\cite{ino10} reproduces the local GLF of flat-spectrum radio quasars (FSRQs) well~\cite{aje12}, although it slightly underestimates the number of FSRQs at $z\gtrsim1$.

In Ref.~\cite{har12} a blazar GLF was constructed based on {\it Fermi} blazar samples following the method of Inoue and Totani~\cite{ino09}, with the constraints using the cumulative source count distribution and the EGB anisotropy measurements \cite{ack12_ani}. Although {\it Fermi} blazar samples were included, the GLF model does not reproduce the observed {\it Fermi} FSRQ GLF even at local redshifts. This may be due to uncertainties in the redshift distribution of observed blazars, which affect parameters of GLFs. However, we note that about a half of {\it Fermi} BL Lacs lack the redshift information~\cite{sha13}. Therefore, it is by no means straightforward to include the redshift information of {\it Fermi} blazars when one attempts to construct a blazar GLF from {\it Fermi} samples.    

In this paper, we utilize GLF model of Ref.~\cite{ino10}. The key parameters of the blazar GLF are $(q,\ \gamma_1, \ \kappa)=(4.50, 1.10,1.42\times10^{-6})$, where $q$ is the ratio of the bolometric jet luminosity and the disk X-ray luminosity, $\gamma_1$ the faint-end slope index of GLF, and $\kappa$ is a normalization factor of GLF.\footnote{See Section. 3 of Ref.~\cite{ino09} for details.} We set minimum and maximum of $L_{\gamma, 100 \rm MeV}$ as 10$^{43}$ erg s$^{-1}$ and 10$^{50}$ erg s$^{-1}$ as in Ref.~\cite{ino10}.

\subsection{Secondary Gamma-ray Spectrum}
\label{subsec:2nd}

\begin{figure}[t]
\begin{center}
\includegraphics[width=0.95\linewidth]{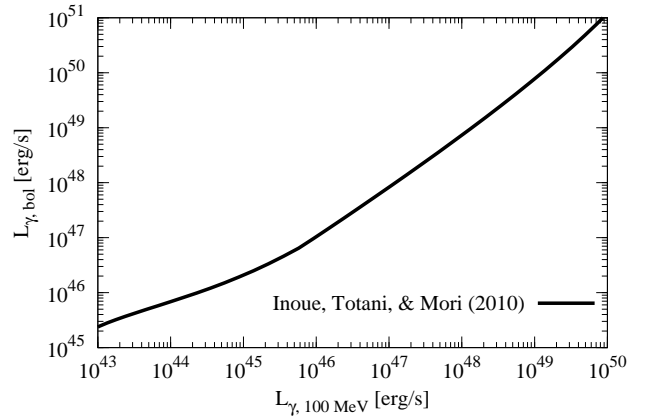}
\caption{Relation between $L_{\gamma, 100 \rm MeV}$ and $L_{\gamma, \rm bol}$ for \cite[][black solid]{ino10}.} 
\label{fig:Lgamma}
\end{center}
\end{figure}
AGNs are believed to produce ultra-high-energy cosmic rays \cite{kot11}. For energies below Greisen--Zatsepin--Kuzmin (GZK) cutoff, pion photoproduction on CMB photons does not occur, and the cosmic rays can propagate cosmological distances without interacting with ambient photons. The deflection of their trajectories depends on the intergalactic magnetic fields (IGMF). The strength of the IGMF is by and large unconstrained.  Theoretical models assuming the dynamo origin of galactic magnetic fields require primordial seed fields of $B>10^{-30}$~G~\cite{Davis:1999bt}, which can be considered as a theoretical lower limit, while the observational upper limit from Faraday rotation is $B<10^{-9}$~G~\cite{Kandus:2010nw,Durrer:2013tn}. Recent gamma-ray observations derived the lower limits that vary from 10$^{-20.5}$~G to 10$^{-15}$~G \cite{ner10,der11,tak13_igmf,ess_magnetic}, depending on the assumptions. From the simulation of large-scale structure formation, the upper limit on the IGMF has been set as $10^{-12}$~G \cite{dol05}. Blazar spectra provide a way to measure IGMFs if the cosmic-ray contribution is taken into account~\cite{ess_magnetic}: the best-fit blazar spectra require IGMFs in the range $(0.01-10)\times 10^{-15}$~G.  The upper limit 
ensures rectilinear propagation of cosmic-ray protons.\footnote{See also Refs.~\cite{mur12,aha13} for discussion of the effects of filaments and clusters.} The lower limit improves the fit to the low-energy part of the blazar spectra.
 In this paper, we consider two representative values of IGMF strengths: $B=10^{-17}$~G and $B=10^{-15}$~G.

Although AGNs are expected to accelerate cosmic rays, the spectrum of cosmic rays produced by AGNs are ambiguous. 
Fortunately, the spectra of secondary gamma rays show almost no sensitivity to variations of the proton injection spectrum~\cite{ess10_prl,ess11}.  
We assume the following form of the proton spectrum~\cite{kal13}:
\begin{equation}
j_p(E)\propto E^{-\alpha}\exp(-E/E_{\rm p,max})\exp(-E_{\rm p,min}/E),
\end{equation}
where we set $\alpha = 2.0$, $E_{\rm p,min}=0.1$~EeV, and $E_{\rm p,max}=1$~EeV. This injection spectrum of protons is among the possibilities that reproduce the blazar spectra~\cite{ess10,ess10_prl}. For the secondary gamma-ray calculation, we use the numerical code~\cite{gel12,kal13}. Our calculation is based on kinetic equations; it calculates the propagation of nucleons, stable leptons and photons using the standard dominant processes, i.e. pion production by nucleons, pair production by protons and neutron $\beta$--decays. For electron-photon cascade development, it includes pair production and IC scattering. 
To model the effects of IGMFs, we mimic deflections in electromagnetic cascades by assigning a finite lifetime to the cascade electrons, assuming that the magnetic field correlation length is always higher than the electrons mean free path (i.e. $l_{\rm cor}>0.1 \ {\rm Mpc}$) and using the angular resolution of CTA with array E \citep{act11}.  While this is a simplification, the results can be tested against the full Monte-Carlo calculations of Essey et al.~\cite{ess_magnetic}, and they agree sufficiently well for our purposes.

Secondary gamma-ray flux depends on the proton luminosity.  Jet power of blazars detected by {\it Fermi} during its first 3-month survey was studied in Ref.~\cite{ghi10}. Based on that study, it is expected that the bolometric proton luminosity, $L_{p,\rm bol}$, is a factor of $\sim 10-100$ larger than the bolometric radiation luminosity, $L_{\gamma,\rm bol}$ for FSRQs, while $L_{p,{\rm bol}}\sim L_{\gamma,\rm bol}$ for BL Lacs. However, it is assumed that electron and positron pairs in jets are negligible. Although pure pair jet models are excluded from X-ray observations of FSRQs~\cite{sik00} and pairs may not survive the annihilation in the inner, compact and dense regions \cite{cel08,ghi10}, there is still room for pairs in the jet, based on the energetics arguments~\cite{sik00,sik05,Boettcher:2013wxa}. In this paper, to be conservative, we assume that $L_{p,{\rm bol}}=L_{\gamma,\rm bol}$. We use the blazar sequence SED to estimate $L_{\gamma, \rm bol}$ from $L_{\gamma, 100 \rm MeV}$ as shown in Figure \ref{fig:Lgamma}.

 In this paper, we adopt the most recent EBL model~\cite{ino13} which constructed EBL based on a semi-analytical galaxy formation model. Inoue et al.~\cite{ino13} successfully reproduce both galaxy evolutionary data and the reionization by taking into account the entire stellar population, including first stars.  
Various experiments have directly observed the EBL at various wavelengths~\cite[e.g.][]{dwe98,fin00,wri00,ede00,mat05,ber07,mat11_akari,mat11,tsu13}. However, direct measurements of the EBL in the optical and near infrared (NIR) bands have been hampered by bright foreground emission caused by interplanetary dust, the so-called zodiacal light~\cite[see][for reviews]{hau01}. Recently, Matsuoka et al.~\cite{mat11} reported measurements of the EBL at 0.44 ${\rm \mu m}$ and 0.65 ${\rm \mu m}$ from outside of the zodiacal region using observational data from {\it Pioneer} 10/11. On the other hand, integration over galaxy number counts provides a firm lower bound on the EBL, and the observed trend of the counts with magnitude indicates that the EBL at $z = 0$ has been largely resolved into discrete sources in the optical/NIR bands~\cite{mad00,tot01,kee10}, even when the effect of incomplete detection due to cosmological dimming of surface brightness is taken into account~\cite{tot01}. Combined with the lower limits from galaxy counts, the total EBL intensity at $z = 0$ from 0.1 ${\rm \mu m}$ to 1000 ${\rm \mu m}$ is inferred to lie in the range 52--99~${\rm nW \ m^{-2} \ sr^{-1}}$~\cite{hor09}. Theoretically, a number of EBL evolution models have been proposed~\cite{mal98,tot02,kne04,pri05,ste06,maz07,fra08,gil09,fin10,kne10,dom11,you11,gil12,hel12,ste12,ino13}. All of these EBL models are in good agreement with the blazar data when secondary gamma rays are included~\cite{ess11}.  The choice of EBL model affects the electromagnetic cascade and changes the implied normalization of the secondary flux in Eqns. (3)-(4) below by a factor of order one.

\section{Results}
\label{sec:res}

\begin{figure*}[t]
\begin{center}
\includegraphics[width=0.95\linewidth]{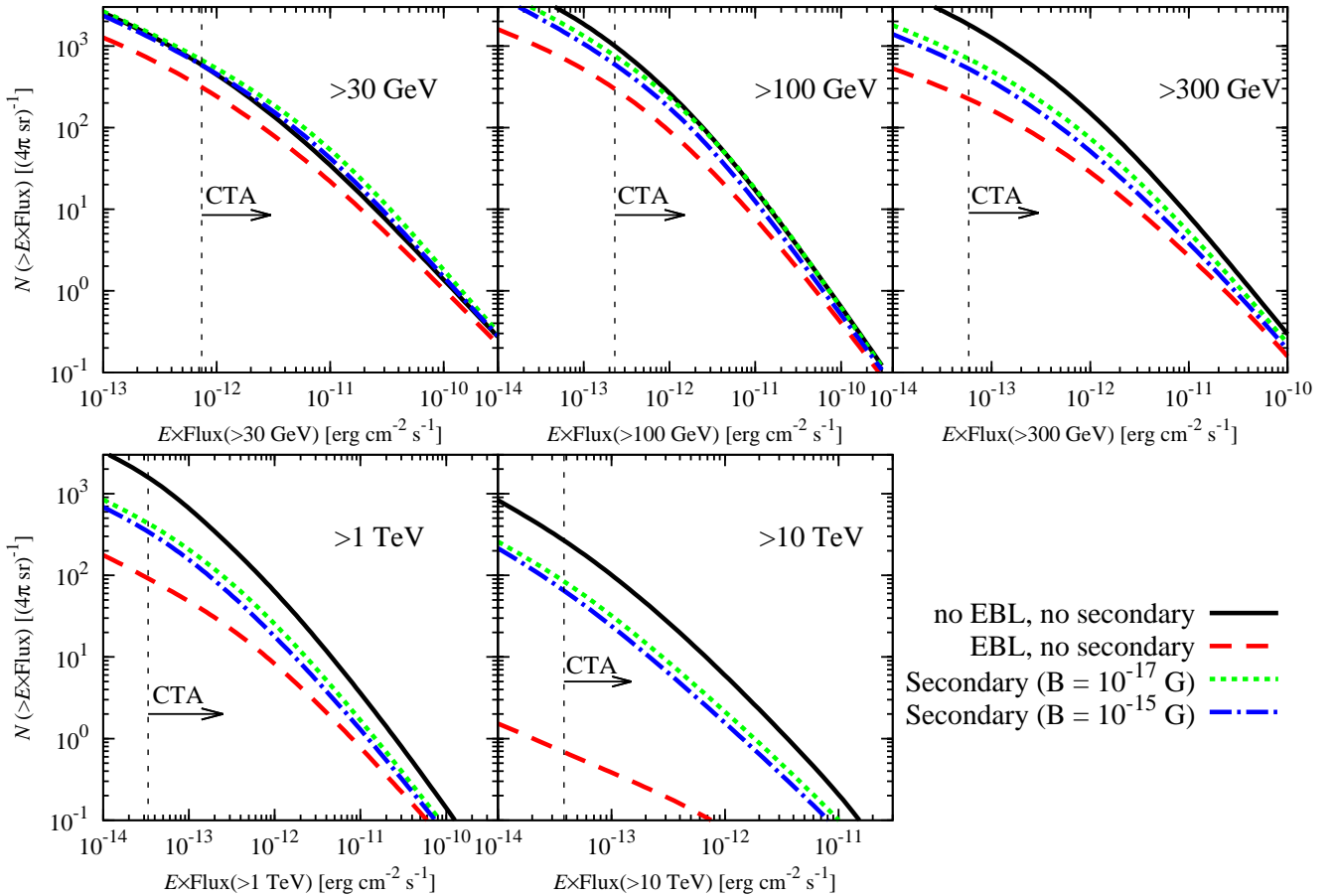}
\caption{Expected cumulative source counts as a function of the integral gamma-ray flux of VHE blazars with a blank field sky survey. The five panels correspond to different photon energies, as indicated in the panels. Solid, dashed, dotted, and dot-dashed curve corresponds to a model without EBL, with EBL, with secondary gamma rays ($B=10^{-17}$~G), and with secondary gamma rays ($B=10^{-15}$~G), respectively. The CTA $5\sigma$, 50 hr detection limit with array E is also shown \cite{dub13}.} 
\label{fig:lognlogs_itm10}
\end{center}
\end{figure*}

\begin{figure*}[t]
\begin{center}
\includegraphics[width=0.95\linewidth]{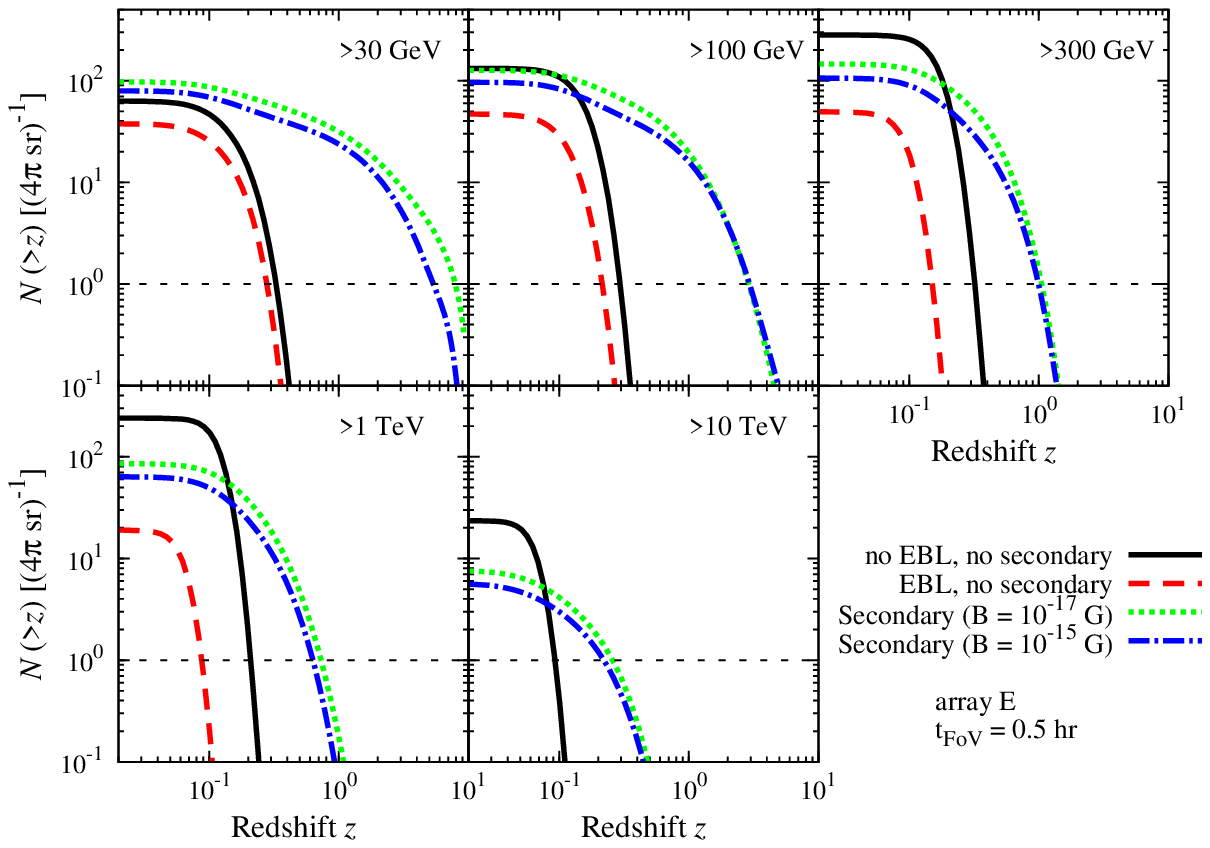}
\caption{Expected cumulative redshift distribution of VHE blazars with a blank field sky survey. The five panels correspond to different photon energies, as indicated in the panels. Solid, dashed, dotted, and dot-dashed curve corresponds to a model without EBL, with EBL, with secondary gamma rays ($B=10^{-17}$~G), and with secondary gamma rays ($B=10^{-15}$~G), respectively. We set the sensitivity of the CTA $5\sigma$, 0.5 hr detection limit with array E \cite{dub13}, where we assume sensitivity limit scales with the inverse square root of observational time. It will take $\sim540$ hr to perform this all sky survey.  The dashed curve in the panel of 10~TeV is not shown, because the expected count is less than 0.1 at overall redshift range.} 
\label{fig:cum_z}
\end{center}
\end{figure*}

A blank field sky survey is the most fundamental mode of observing the sky in an energy band, free from any preselection biases,  except for the flux limit of the survey. For CTA, various survey designs are possible for a fixed amount of the total observation  time, changing the survey area and exposure time for each field of view (FoV) \cite[e.g.][]{ino10,dub13}. The dependence of source counts on the key parameters of the survey -- FoV, $\theta_{\rm FoV}$, observing time per FoV, $t_{\rm FoV}$, total observing time, $t_{\rm obs}$ -- can be estimated analytically~\cite{dub13}. The total source count $N[>F(t_{\rm FoV})]$ above a certain flux limit $F$ is
\begin{equation}
N[>F(t_{\rm FoV})]\propto A_{\rm obs}F(t_{\rm FoV})^{-n}\propto t_{\rm obs} \theta_{\rm FoV}^2 t_{\rm FoV}^{-(1-n/2)},
\end{equation}
where $A_{\rm obs}$ is the  total survey area with a fixed time $t_{\rm obs}$ and $n$ is the slope index of the cumulative source count distribution. We assume that flux limit scales as the inverse square root of observation time.  The flux of primary gamma-rays scales with 
distance $d$ as
\begin{equation}
F_{\gamma, \rm primary} \propto d^{-2} \exp(-d/\lambda_{\gamma\gamma}),
\label{eq:horizon_prim}
\end{equation}
where $\lambda_{\gamma\gamma}$ is mean free path of pair production on 
EBL. Neglecting evolution and pair production on EBL one can obtain $n$ 
is 1.5 in the Euclidean universe because $N\propto d^3\propto F^{-3/2}$.  Then, $N[>F(t_{\rm FoV})]\propto t_{\rm obs} \theta_{\rm FoV}^2 
t_{\rm FoV}^{-0.25}$. Due to EBL attenuation effect as well as the 
cosmic expansion and evolution of sources $n$ generally becomes smaller 
at fainter flux. Therefore, a wide and shallow survey is favored to 
detect more sources.

Secondary gamma rays change this relation. In contrast 
to~(\ref{eq:horizon_prim}) the flux of secondary gamma rays with distance $d$ 
is approximated as \cite{ess11,ess12}
\begin{eqnarray}
F_{\gamma, \rm secondary}
 &\propto \left\{\begin{array}{ll}
	d^{-1} &  (d \ll \lambda_{\gamma\gamma}), \\
	d^{-2} & (d \gg   \lambda_{\gamma\gamma}),
    \end{array}\right.
    \label{eq:horizon}
\end{eqnarray}
Then, $n$ becomes 3.0 at $d \ll \lambda_{\gamma\gamma}$ and $n=1.5$ at 
$d \gg \lambda_{\gamma\gamma}$.
In the first case $N[>F(t_{\rm FoV})]$ will be proportional to $t_{\rm 
obs} \theta_{\rm FoV}^2 t_{\rm FoV}^{0.5}$. This suggests a narrow and 
deep survey is favored to detect more sources. However, we note that $n$ 
may be smaller than 3 due to the cosmic expansion and the evolution 
effects. In any case we expect more sources assuming the secondary 
component dominates the gamma-ray flux, although flux from nearby 
sources is dominated by the primary flux.

Figure \ref{fig:lognlogs_itm10} shows the cumulative source count distributions in the entire sky above five energy thresholds (30 GeV, 100 GeV, 300 GeV, 1 TeV, and 10 TeV).  The CTA $5\sigma$, 50 hr detection limit with array E is also shown \cite{dub13}. Absorption of blazar spectra by the EBL is taken into account using \cite{ino13} and secondary gamma rays is also taken into account using \cite{kal13}. When we do not take into account secondary gamma rays, the expected source counts are rapidly flattened at fainter flux and decrease at higher energy band due to the EBL attenuation effects.  This is in agreement with results of Ref.~\cite{dub13}. 

Let us consider an all sky survey with 50 hr exposures of each FoV. When we consider the effect of the EBL attenuation only, the expected number of blazars are 350, 340, 240, 100, and $<1$ above 30~GeV, 100~GeV, 300~GeV, 1~TeV, and 10~TeV, respectively. Once we take into account secondary gamma rays, these numbers will increase as 720, 840, 760, 460, and 93, and 630, 650, 530, 360, and 72 for the case of $B=10^{-17}$~G and $B=10^{-15}$~G, respectively. Due to the magnetic deflection, the expected number of blazars is highly affected by the magnetic field strength especially at $\lesssim1$ TeV. It will be difficult to explain such a large number of blazars at $>1$ TeV by the evolution or the EBL modeling. Evolution at each energy band should be similar and the gamma-ray horizon where gamma-ray opacity becomes unity is typically at $z\sim0.15$ and $\sim0.02$ at 1 and 10 TeV, respectively \cite{ino10}.  If secondary gamma rays do not contribute significantly to VHE gamma-ray emission from blazars, there should be a large drop in blazar counts at higher energy. Therefore, cumulative source count distribution will be a clear evidence of the secondary gamma rays. However, it requires $\sim 5.4\times10^4$ hr to survey all sky with exposures of 50 hr at each FoV assuming the FoV of CTA as $7^\circ$ \cite{dub13}. Such an exposure is unrealistic for a wide survey limited in time.

Let us consider an all sky survey with a fixed observing time by multiple CTA pointing observations assuming the FoV of CTA to be $7^\circ \sim 40 \ {\rm deg}^2$ and considering array E configuration~\cite{act11,dub13}. A 0.5 hr exposure of each FoV will allow one to survey all sky ($4\pi$~str) in $\sim540$ hr (i.e. $\sim1000$ pointing observations). Since the typical total observable time for IACTs is 1000 hr in a year, and CTA will have both north and south sites to cover the entire sky, this long term observation can be accomplished as a multi-year project. We note that the detection limit of 0.5~hr exposure of CTA is equivalent to that of 50~hr exposure of current generation of IACTs.

Figure \ref{fig:cum_z} shows the cumulative redshift distribution in the entire sky above five energy thresholds  (30 GeV, 100 GeV, 300 GeV, 1 TeV, and 10 TeV) with 0.5 hr exposure for each FoV. We estimate the 5$\sigma$ sensitivity limit with 0.5~hr exposure from $5\sigma$, 50~hr detection limit by assuming that sensitivity limit scales as the inverse square root of observation time~\cite{dub13}. For $B=10^{-17}$~G, the secondary gamma-ray scenario predicts 100, 130, 150, 87, and 8 blazars above 30 GeV, 100 GeV, 300 GeV, 1 TeV, and 10 TeV, respectively. For $B=10^{-15}$~G, these numbers become 79, 96, 110, 63, and 6, respectively. We summarize the expected blazar counts for this survey mode in Table~\ref{tab:count_0.5hr}. The highest redshift will extend up to $z\sim8$ and $z\sim5$ for $B=10^{-17}$ and $10^{-15}$~G, respectively.  This prediction differs dramatically from 
what would be expected in the absence of secondary gamma rays, in which case the highest redshift is $z< 0.4$. As discussed above, this difference 
arises because secondary gamma rays are generated not at the source, but continuously along the line of sight, and the fraction produced 
relatively close to the observer is not attenuated. Since stronger IGMF suppress the secondary gamma-ray flux at lower energy, lower IGMF allows us to detect higher redshift objects. Therefore, statistical samples with a shallow survey will allow one to obtain 
a clear evidence of the secondary gamma-ray production and to probe the IGMF strength.

\begin{table*}
 \begin{center}
  \caption{Expected blazar counts for all sky survey with 0.5 hr exposure of each FoV}
  \label{tab:count_0.5hr}
  %   \begin{minipage}{80mm}
   \begin{tabular}{lccccc}
      \hline
      &   \multicolumn{5}{c}{Energy} \\ \cline{2-6}
      & $>$30 GeV & $>$100 GeV &  $>$300 GeV & $>$1 TeV & $>$10 TeV \\ \hline
		no EBL, no secondary 					& 63 	& 130 	& 280 	& 240 	& 24 \\
		EBL, no secondary 					& 38 	& 47 	& 50 	& 20 	& $<1$ \\
		Secondary ($B=10^{-17}$ G) & 100 	& 130 	& 150 	& 87 	& 8\\
		Secondary ($B=10^{-15}$ G)  & 79 	& 96 	& 110 	& 63 	& 6 \\ \hline
    \end{tabular}
%   \end{minipage}
  \end{center}
\end{table*}

Let us now consider 50~hr exposure of each FoV with total observing time $\sim540$ hr (i.e. 10--11 pointing observations). One can cover only $\sim1$\% of the entire sky in this mode. The secondary gamma-ray scenario predicts the expected number of sources as 7, 8, 8, 5, and 1 above 30~GeV, 100~GeV, 300~GeV, 1~TeV, and 10~TeV, respectively, in the case of $B=10^{-17}$~G, while these numbers will be 6, 7, 5, 4, and 1 in the case of $B=10^{-15}$~G. Therefore, in a fixed observable time, a wider and shallower survey will have an advantage in being able to detect more sources in total, and more sources at higher redshifts, in all energy bands.

In this paper, we assume $L_{\rm p,bol}=L_{\gamma,\rm bol}$ (see Section. \ref{subsec:2nd} for details). However, the ratio of proton and photon powers is still unknown. If proton power is ten times higher than photon power,  $L_{\rm p,bol}=10L_{\gamma,\rm bol}$, the expected counts for the all sky survey mode with 0.5 hr exposure for each FoV will increase. Those will be 490, 620, 690, 440, and 92 blazars above 30 GeV, 100 GeV, 300 GeV, 1 TeV, and 10 TeV, respectively, for $B=10^{-17}$~G, while those will be 320, 400, 440, 330, and 72, respectively, for $B=10^{-15}$~G. These counts are not simply scaled by a factor of 10 from those in the case of $L_{\rm p,bol}=L_{\gamma,\rm bol}$ due to the cosmological evolution of blazars.

After one determines the spectral template of primary gamma rays from blazars and the blazar GLF at an energy band where EBL and secondary gamma rays do not alter the spectra of sources, cumulative source count distribution at VHE  will enable one to constrain the strength of the IGMF by comparing the observed counts with the expected counts. Although one must assume a specific EBL model, EBL models are currently well constrained by various galaxy formation and reionization data \cite[see e.g.][]{ino13}.

We note that there are uncertainties in the numbers predicted above. The use of the blazar SED sequence and the secondary gamma-ray SED is the key to converting the blazar luminosity function in the GeV band into the VHE band. However, the validity of the blazar sequence is still a matter of debate \cite[e.g.][]{pad07} and secondary gamma ray flux is strongly affected by the strength of IGMF along the average line of sight, which is unknown. 
We did not consider the time variability of blazars, which is expected to occur on different time scales for primary and secondary gamma rays~\cite{Prosekin:2012ne}. The luminosity function model parameters of Ref.~\cite{ino10} have been determined only by about 50 EGRET blazars, although it successfully reproduced {\it Fermi} FSRQ local GLF. Since a half of {\it Fermi} BL Lacs have unknown redshifts~\cite{sha13}, it is still not easy to evaluate the blazar GLF including BL Lacs with current {\it Fermi} samples. A blazar GLF with future complete {\it Fermi} blazar sample will enable us to make more precise prediction. Moreover, our GLF is uncertain for $z > 3$ because the current observed number of X-ray AGNs and gamma-ray blazars above $z \sim 3$ is insufficient to strongly constrain the model \citep{ino11_highz} \footnote{A gamma-ray blazar candidate at $z\sim3$--4 is reported \citep{tak13_blz}.}. Finally, our model includes only the known blazar population. A completely different and new extragalactic population such as starburst galaxies and radio galaxies may be found by a blind survey, which is probably the most exciting possibility and a strong motivation for the survey.

\section{Summary}
\label{sec:sum}
We have considered the impact of secondary gamma rays~\cite{ess10,ess11,ess12,Prosekin:2012ne,mur12,Razzaque:2011jc,aha13,kal13} on the future of VHE gamma-ray sky survey by CTA. We find that secondary gamma rays will enable CTA to detect 100, 130, 150, 87, and 8 blazars above 30 GeV, 100 GeV, 300 GeV, 1 TeV, and 10 TeV, respectively, in the case of 0.01 femtogauss strength ($B=10^{-17}$~G), while the same numbers will be 79, 96, 110, 63, and 6 in the case of IGMFs of femtogauss strength ($B=10^{-15}$~G). The required total survey time will be 540~hr, where we set 0.5 hr exposure for each FoV. In the absence of secondary gamma rays, one would not expect such a large number of sources~\cite{ino10,dub13}. Moreover, secondary gamma rays will enable CTA to observe sources up to $z\sim5$--8. To observe a large number of sources, a wider and shallower survey is favored. The enhancement of the number of sources and wide redshift range of blazars in a future CTA blazar samples obtained by the survey will elucidate the role of secondary gamma-ray scenario and enable statistical studies of the properties of blazars, such as their evolution.  

The detectable number of blazars is also sensitive to IGMFs due to magnetic deflection effect which is significant at lower energy. Redshift distribution of the CTA blazars obtained by the survey enables us to probe IGMFs statistically. This method will be a complementary method to constrain the IGMF strength from spectra of individual sources.

\subsubsection*{Acknowledgements:} 
The authors thank Warren Essey and Kohta Murase for useful comments and discussions. YI acknowledges support by the Research Fellowship of the Japan Society for the Promotion of Science (JSPS). O.E.K. was supported by the grant of the Russian Ministry of Education and Science No. 8412, the grant of the President of the Russian Federation NS-5590.2012.2, and the grant of RFBR 13-02-01293.  A.K. was supported by the Department of Energy Award Number DE-SC0009937 and by the World Premier International Research Center Initiative (WPI Initiative), MEXT, Japan.  A.K. appreciates the hospitality of the Aspen Center for Physics, which is supported by the NSF Grant No. PHY-1066293.

%% References


\bibliographystyle{model1a-num-names}
% \bibliographystyle{elsarticle-num}






% \end{thebibliography}

\end{document}